\documentclass[12pt]{article}
\usepackage{graphicx} 
\usepackage{subfigure}

\def\be{\begin{equation}}
\def\ee{\end{equation}}
\def\bea{\begin{eqnarray}} \def\eea{\end{eqnarray}} \def\lb{\label}
\newcommand{\half}{\mbox{$\frac{1}{2}$}}
\newcommand{\third}{\mbox{$\frac{1}{3}$}} \newcommand{\e}{{\rm e}}

\begin{document}

\begin{center} {\bf \Large Ising Models for Inferring Network Structure
From Spike Data}

\vspace*{0.5cm} {\large J. A. Hertz  \\} {\small Niels Bohr Institute,
Blegdamsvej 17, 2100 Copenhagen {\O}, Denmark\\ NORDITA,
Roslagstullsbacken 23, 10691 Stockholm, Sweden}

\vspace*{0.2cm} {\large Yasser Roudi \\} {\small Kavli Institute for
Systems Neuroscience, NTNU, Trondheim, Norway\\ NORDITA,
Roslagstullsbacken 23, 10691 Stockholm, Sweden}

\vspace*{0.2cm} {\large Joanna Tyrcha\\} {\small Mathematical
Statistics, Stockholm University, S106 91 Stockholm, Sweden}
\end{center}

\vspace*{0.5cm} \subsection*{Abstract} Now that spike trains from many
neurons can be recorded simultaneously, there is a need for methods to
decode these data to learn about the networks that these neurons are
part of.  One approach to this problem is to adjust the parameters of a
simple model network to make its spike trains resemble the data as much
as possible.  The connections in the model network can then give us an
idea of how the real neurons that generated the data are connected and
how they influence each other.  In this chapter we describe how to do
this for the simplest kind of model: an Ising network.  We derive
algorithms for finding the best model connection strengths for fitting a
given data set, as well as faster approximate algorithms based on mean
field theory.  We test the performance of these algorithms on data from
model networks and experiments.

\newpage

\noindent {\bf \large 1 Introduction} \vspace{0.2 cm}

Now that we can record the spike trains of large numbers of neurons
simultaneously, we have a chance, for the first time in the history of
neuroscience, to start to understand how networks of neurons work. But
how are we to proceed, once we have such data?  In this chapter, we will
review some ideas we have been developing.  The reader will recognize
that we are only describing the very first steps in a long journey.  But
we hope that they will help point the way toward real progress some time
in the not-too-distant future.

What is it we want to learn from multi-spiketrain data? Here we will
focus on finding the connectivity in the network under study.  We hope
that ultimately this can help us understand the dynamics of the network
and how it implements computations.   For example, in the hippocampus
and entorhinal cortex, many interesting kinds of neurons (place cells,
grid cells, etc.) have been identified, but we do not know how they
interact.  If we could find out how they are connected, we might be able
to get a much better idea of how they compute the animal's estimate of
its location.

Of course, we will not be able to find the full connectivity unless we
can record from {\em all} neurons, which is impossible, at least with
any current techniques.  Several orders of magnitude separate the number
of electrodes in a recording array and the number of neuron in the area
it covers, so any apparent effect of one neuron on another might
actually be mediated by unobserved neurons. Thus, we (and everyone else
in the field) are forced into using terms like ``effective connections''
or ``functional connectivity.''  In this chapter, for brevity, we will
generally just write ``connections" or ``connectivity", but we will
always mean that they are ``effective'' or ``functional'' connections.

Given this fundamental limitation on our project, it is nevertheless
worth remarking that while all connections are functional one, some
kinds are more functional than others.  Modeling approaches that are
closer to biophysical reality stand a better chance of recovering
something like the real synaptic connectivity than those which are
farther away.  The approach we describe here is at least a first step,
albeit a modest one, in the right direction.

One can take a rather different attitude, proceeding formally and making
a purely statistical model: a mathematical object with some parameters
that characterize the data.  However, even if such a model fits the data
well, one has in general no way to assign biological meaning to the
parameters.  This is not to say that we don't care about fit quality --
we do.  It is only to say that fitting, for example, equal-time neuronal
firing correlations in the data perfectly with some model is not very
satisfying if the parameters of the model cannot be related fairly
directly to the connectivity of the network.

To present the models and formal approaches that we study, we have to
start with our notation for the data.  We work with time-binned spike
trains under the assumption that firing rates are low enough that there
is (almost) never more than one spike per bin.  Time will be measured in
units of the bin size.  We denote the spike train of neuron $i$ by
$S_i(t)$, where $S_i(t)=+1$ if it fires in bin $t$ and $S_i(t)=-1$ if it
does not. (One can equivalently use $S_i = 0$ for no spike, but we won't
do that here.)  Thus we can visualize the data as a big array, the
``spike matrix'' ($N \times T$ if there are $N$ neurons and $T$ time
bins), of $+1$s and $-1$s.

This representation of the data lends itself to formulating the problem
in terms of a very simple, perhaps the simplest possible model: a
McCulloch-Pitts network (McCulloch and Pitts, 1943), or what in physics
is called an Ising model.  In this chapter we will use several different
kinds of Ising models to treat the problem of finding the connections in
a network from its spiking history.

\vspace{0.75 cm} \noindent {\bf \large 2 The Gibbs Equilibrium Model}
\vspace{0.2 cm}

A possible formal approach to the problem is this (Schneidman et al,
2006):  One considers the data as the set of columns on the spike
matrix, i.e., all the {\em spike patterns} observed, ignoring their
temporal order, and asks what distribution they could have been sampled
from. Since one is treating all the patterns as coming from the same
distribution, one is implicitly assuming stationarity in the data.  If
one seeks the distribution with the largest entropy, consistent with the
sample means $m_i = \langle S_i\rangle$ and correlations $\langle
(S_i-m_i)(S_j-m_j)\rangle$,\footnote{Statisticians would call these
covariances, but here we follow the convention in statistical physics
and call them correlations.} one finds something with a familiar look
(at least if one is a statistical physicist): \be P[{\bf S}] =
\frac{1}{Z}\exp\left[ \half \sum_{ij}J_{ij}S_iS_j + \sum_i h_iS_i\right]
\lb{Ising} \ee This is the Gibbs equilibrium distribution for the
(pairwise) Ising model.  Its parameters, $J_{ij}$ and $h_i$, are
Lagrange multipliers that are used in the constrained maximization.  The
parameter $J_{ij}$ is the strength with which unit (neuron) $j$
influences unit $i$, and the bias ``field'' $h_i$ controls how likely
unit $i$ is to be +1 (``fire'') in the absence of the other units.   The
expectation values of the $S_i$ under the distribution (\ref{Ising}),
denoted $m_i$ (``magnetizations'' in the original context of the model),
are related to the neuronal firing rates $r_i$ (in units of spikes/time
bin) through  $m_i = 2r_i - 1$.

One might like to think of $J_{ij}$ as something like a synaptic
strength.  However, the $\sf J$ matrix in (\ref{Ising}) is necessarily
symmetric, $J_{ij} = J_{ji}$. This is true for the couplings in magnets,
but it is generally not true for synapses, so we have to be cautious
about what $J_{ij}$ means.

A few remarks:  In physics, the normalizing denominator $Z$ in
(\ref{Ising}) is called the partition function, and its log is the
negative of the free energy.  We are everywhere setting the temperature
equal to 1, since changing the temperature just amounts to rescaling the
$J_{ij}$s and $h_i$s by a constant factor.

Normally in statistical mechanics, one is given a model and its
parameters, and the task is to find the moments $\langle S_i \rangle$,
$\langle S_iS_j \rangle$, etc., which are the quantities that can be
measured in experiments. But here  we have an {\em inverse problem}: we
are given the measured correlation functions and want to find the
parameters of the model.

It has been known for some time how to solve this problem, at least in
principle.  One should adjust the $h_i$ and $J_{ij}$ to maximize the
probability (\ref{Ising}), evaluated on the data.  Doing this by
gradient ascent gives learning rules \bea \delta h_i &=&   \eta (
\langle S_i  \rangle_{\rm data} - \langle S_i  \rangle_{{\sf J}, {\bf
h}}), \lb{Boltzh}		\\ \delta J_{ij} &=& \eta  ( \langle S_i S_j
\rangle_{\rm data} - \langle S_i  S_j\rangle_{{\sf J}, {\bf h}}).
\lb{BoltzJ} \eea The averages in the second terms are under the model
with the current values of the $J_{ij}$ and $h_i$, and $\eta$ is a
learning rate, to be chosen small enough to get smooth convergence.  To
evaluate them exactly, we have to know the probabilities (\ref{Ising}),
which requires evaluating the partition function $Z$.  To do this
exactly one has to sum $2^N$ terms, which is feasible for systems up to
$N \approx 20$.  For larger $N$ one has to estimate the averages by
Monte Carlo simulations of the model.   This fact makes the learning
slow, especially when working with data from long recordings.  One wants
the estimates of the model averages to be as good as the direct data
averages in the first terms, so the length of each Monte Carlo run has
to be about equal to the number of time bins in the data set.  It may
take many iterations to reach stationary values of the $J_{ij}$ and
$h_i$, so this can be a lot of computing.  Just what limit this places
on the size of systems that can usefully be studied this way depends on
patience and the computing resources available, but in our work we found
it impractical to try to work with $N > 100$.

These learning rules are a special case of what is called Boltzmann
learning (Ackley et al, 1985), which also covers systems with
``hidden'', i.e., unrecorded units.  In this case, the first terms on
the right-hand sides of (\ref{Boltzh}) and (\ref{BoltzJ}), when
evaluated for hidden units, are averages over the model with the visible
units clamped to their measured values.  This requires even more Monte
Carlo runs.

\vspace{0.5 cm} \noindent {\bf 2.1 Mean-field methods } \vspace{0.2 cm}

It is possible to avoid long Monte Carlo runs using mean field methods. 
Two such methods have been proposed.  One is based on heuristic
mean-field equations first applied to magnetic systems by Weiss more
than a century ago.  The other is based on improved equations first
proposed  by Onsager.   They were applied to spin glasses by Thouless,
Anderson and Palmer (Thouless et al, 1977), so nowadays in statistical
mechanics they are called TAP equations.

Both the original mean-field and TAP equations are approximate partial
solutions to the forward problem, i.e., calculating the magnetizations
$m_i$.  From them, we will derive corresponding solutions to the inverse
problem. These solutions are approximate, but become very good in the
limits of weak coupling or, for dense connections, large $N$.

The idea of mean field theory is very simple.  The exact $m_i$,
conditional on a set of $S_j$ connected to $i$ by the coupling matrix
$J_{ij}$, is the difference of Boltzmann probabilities \bea p(S_i=1|
\{S_j\}) - p(S_i=-1| \{S_j\}) &=&  \frac{{\e}^{h_i +
\sum_jJ_{ij}S_j}-{\e}^{-h_i - \sum_jJ_{ij}S_j}}{{\e}^{h_i +
\sum_jJ_{ij}S_j}+{\e}^{-h_i - \sum_jJ_{ij}S_j}} \nonumber \\ &=&  \tanh
\left( h_i + \sum_jJ_{ij}S_j\right). \lb{mexact} \eea The approximation
consists of replacing the $S_j$  inside the tanh by their averages
$m_i$: \be m_i = \tanh \left( h_i + \sum_jJ_{ij}m_j\right). \lb{MFT} \ee
This is apparently a good approximation when the sum on $j$ has many
terms (a loose kind of central-limit argument).  This is called the mean
field limit.

Onsager argued that the contribution to the neighbor magnetizations
$m_j$ from the central unit $S_i$ itself should not be counted in
estimating the field acting on $S_i$.  This leads to corrected
mean-field equations, \be m_i = \tanh \left[ h_i + \sum_jJ_{ij}m_j
-m_i\sum_jJ_{ij}^2(1-m_j^2)\right]. \lb{TAP} \ee TAP showed that these
equations, rather than (\ref{MFT}), should be used in spin glasses,
where the $J_{ij}$ are random, with a zero or very small mean, because
then the Onsager correction term is of the same order as the naive mean
field.  In this sense, these equations are the correct mean field
equations for spin glasses. It has become conventional to call them
``TAP equations''.  We will use the abbreviation ``nMF'' for the naive
mean field equations (\ref{MFT}).

Plefka showed that (\ref{MFT}) and (\ref{TAP}) are the first two in a
sequence of better and better approximations that can be derived
systematically (Plefka, 1982), but we will only consider these first two
here.

It is convenient to write the TAP equations in the form \be h_i =
\tanh^{-1}m_i - \sum_jJ_{ij}m_j + m_i\sum_jJ_{ij}^2(1-m_j^2)
\lb{TAPhofm} \ee The matrix \be \chi^{-1}_{ij} = \frac{\partial
h_i}{\partial m_j} = \frac{\delta_{ij}}{1-m_i^2} -J_{ij}
-2J_{ij}^2m_im_j \lb{TAPchiinv} \ee is the inverse susceptibility
matrix. In equilibrium statistical mechanics there is a theorem that the
susceptibility matrix is (up to a factor of the temperature, which we
are setting equal to 1) equal to the correlation matrix \be C_{ij} =
\langle (S_i - m_i)(S_j - m_j)\rangle \lb{correlations} \ee Thus, if we
know the correlation matrix, we simply need to invert it and solve (for
$i\neq j$) \be ({\sf C}^{-1})_{ij} = -J_{ij} -2J_{ij}^2m_im_j
\lb{inversionformula} \ee for $J_{ij}$ (Kappen and Rodriguez, 1998;
Tanaka, 1998; Roudi et al, 2009).  This is the TAP algorithm.  There are
$n(n-1)/2$ quadratic equations to be solved, but they are decoupled. 
For the original, ``naive'' mean field approximation, it is even
simpler: The last term on the right hand side of
(\ref{inversionformula}) is absent, and we just have $J_{ij} = -({\sf
C}^{-1})_{ij}$.

\vspace{0.75 cm} \noindent {\bf \large 3 Kinetic Ising Models  }
\vspace{0.2 cm}

The Ising model as described so far has no dynamics.  It is defined
solely by the Gibbs equilibrium distribution (\ref{Ising}).  Since the
system we are trying to understand is a noisy dynamical one, we would
like to fit its data -- the spike trains -- by a stochastic dynamical
model.  And the Ising model can be given a dynamics in a natural way, as
follows (Glauber, 1963).  At each time, every neuron has a chance
$\propto dt$ of being updated in the infinitesimal interval $[t,t+dt)$
(i.e., neuron updates are independent Poisson processes).  For a neuron
that is updated, we compute the total ``field'' \be H_i(t) = h_i(t)
+\sum_jJ_{ij}S_j(t). \lb{fieldatt} \ee (We can allow the external field
$h_i$ to depend on time.)  We then let $S_i$ take its next value
$S_i(t+\Delta t)$ according to the probability \be P(S_i(t+\Delta t) |
\{ S_j(t) \}) = \frac{\exp [S_i(t+\Delta t)H_i(t)]}{2 \cosh H_i(t)}=
\half [ 1 - S_i(t+\Delta t) \tanh H_i(t)]. \lb{KIdef} \ee It is then
possible to show that if $h_i$ is independent of $t$ and the matrix $\sf
J$ is symmetric, this model has a stationary distribution given by
(\ref{Ising}).

Since we would like to identify the $J_{ij}$ with synaptic strengths,
which are not symmetric, we relax the symmetry condition.  Furthermore,
here we will update all the neurons simultaneously instead of randomly
asynchronous, because it makes the model easier to apply to time-binned
data, such as we are assuming we have here.  (We set $\Delta t = 1$ from
now on.) With either of these changes, the Gibbs equilibrium
distribution (\ref{Ising}) is no longer a stationary solution of the
dynamics, even if the $h_i$ are time-independent.  In the case that they
are and $\sf J$ is symmetric, there is a stationary distribution, \be
P[{\bf S}] = \frac{\prod_i [2\cosh (\sum_jJ_{ij}S_j + h_i)]
}{\sum_{\{\sigma \}}\prod_i [2\cosh (\sum_jJ_{ij}\sigma_j + h_i)] }
\lb{synchequildist} \ee (Peretto, 1984). When the $h_i$ are
time-independent and $\sf J$ is not symmetric, a stationary distribution
may still exist, but it cannot be written down in closed from.  Since
such a state is not described by the Gibbs equilibrium distribution
(\ref{Ising}), we call it a non-equilibrium state, even though it is
stationary.
\newpage
\vspace{0.5 cm} \noindent {\bf 3.1  Stationary case: exact algorithm }
\vspace{0.2 cm}

For simplicity, we consider first the case where the $h_i$ are
time-independent and we assume a stationary distribution of the $S_i$.
If we are given a set of spike trains in the form ${\bf S} =\{ S_i(t)
\}$, $1 \le i \le N$, we can derive an algorithm for finding the model
parameters $J_{ij}$ and $h_i$ by performing gradient ascent on the
log-likelihood of the data under the model, which is \be L[{\bf S}, {\sf
J}, {\bf h}] = \sum_{it} [S_i(t+1)H_i(t) - \log 2 \cosh H_i(t)], \lb{LL}
\ee with $H_i(t)$ given by (\ref{fieldatt}).  This gives learning rules
\bea \delta h_i  &=& \eta \left [ \langle S_i(t+1) \rangle_t  - \left
\langle \tanh H_i(t) \right\rangle_t \right]     \lb{dh}\\ \delta J_{ij}
&=& \eta \left [  \langle S_i(t+1)S_j(t) \rangle_t - \left \langle \tanh
H_i(t) S_j(t)\right \rangle_t	\right]		\lb{dJ} \eea (Roudi and
Hertz, 2011a). Equations (\ref{dh}) and (\ref{dJ}) have a form analogous
to (\ref{Boltzh}) and (\ref{BoltzJ}) for the equilibrium case: the right
hand sides are differences between averages over the data  and averages
under the current model.  However, here the latter  can be evaluated
directly and quickly from the data, so this algorithm is generally much
faster than the equilibrium one.

It is practical in implementing this algorithm to express the neuronal
firing variables in terms of the differences $\delta S_i(t) = S_i(t) -
m_i$, with $m_i = \langle S_i(t) \rangle_t$ and to write $H_i(t)$ in the
form \be H_i(t) = b_i + \sum_j J_{ij} \delta
S_j(t)			\lb{newHdef} \ee with \be b_i = h_i + \sum_j J_{ij}
m_j. \ee Then we have \bea \delta b_i  &=& \eta \left [ m_i  - \left
\langle \tanh H_i(t) \right\rangle_t \right]     \lb{db}\\ \delta J_{ij}
&=& \eta \left \{ D_{ij} - \left \langle \left [\tanh  H_i(t) - m_i
\right ] \delta S_j(t)\right \rangle_t	\right\},	\lb{dJ2} \eea
with \be D_{ij} = \langle \delta S_i(t+1) \delta S_j(t)
\rangle,			\lb{Ddef} \ee the one-step-delayed correlation
matrix.  The factor in brackets in the time average in (\ref{dJ2}) can
be recognized as the expectation value of $\delta S_i(t+1)$, given the 
measured $S_j(t)$, so the second term is the one-time-step delayed
correlation matrix for the model, conditional on the data at $t$.

This algorithm is exact, in the sense that for data generated by a
kinetic Ising model of this form, it will recover the $J_{ij}$ and $h_i$
exactly, after infinite iterations, for infinite data.  (By ``infinite
data'' we mean an infinite number of time steps.)

\vspace{0.5 cm} \noindent {\bf 3.2  Stationary case: mean-field methods
} \vspace{0.2 cm}

Like the equilibrium model, this model allows systematic and rapid
mean-field and TAP approximations (Roudi and Hertz, 2011a).  To derive
the mean-field algorithm, we start by assuming that the $m_i$ (of the
data) satisfy the mean-field equations (\ref{MFT}).  Then, on the
right-hand side of (\ref{dJ}), we write $S_i(t) = m_i + \delta S_i(t)$
and expand the tanh to first order in the $J_{ik}$.  First-order terms
in $\delta S$ cancel, and we are left with \be \delta J_{ij} = \langle
\delta S_i(t+1) \delta S_j(t) \rangle - (1-m_i^2)\sum_k J_{ik} \langle
\delta S_k(t) \delta S_j(t) \rangle .				\lb{MF1} \ee If the
$J_{ik}$ are the true ones, then we should have $\delta J_{ij}=0$.  Thus
we obtain a simple matrix equation \be {\sf D} = {\sf A}{\sf J}{\sf
C}				\lb{MF2} \ee where \bea C_{ij}&=& \langle \delta
S_i(t) \delta S_j(t) \rangle				\lb{Cdef} \\ A_{ij} &=&
(1-m_i^2)\delta_{ij}					\lb{Adef} \eea The $J$ matrix
can be found simply by matrix inversion and multiplication in terms of
first- and second-order statistics of the data: \be {\sf J}  = {\sf
A}^{-1}{\sf D}{\sf C}^{-1}.				\lb{MFJ} \ee Knowing the
$J_{ij}$, the inversion task is then completed by solving the MF
equations (\ref{MFT}) for $h_i$: \be h_i = \tanh^{-1}m_i - \sum_j
J_{ij}m_j. \ee

Likewise, we can get a TAP approximation by assuming that the $m_i$ of
the data satisfy the TAP equations (\ref{TAP}) and expanding the tanh to
third order.  After a little algebra, and ignoring some terms which are
small for networks with weak, dense connections, we again find the
formula (\ref{MF2}), but where now \be A_{ij} =
(1-m_i^2)(1-F_i)\delta_{ij},		\lb{ATAP} \ee with \be F_i =
(1-m_i^2)\sum_k J_{ik}^2 (1-m_k^2).		  \lb{Fdef} \ee We cannot
solve for the $J_{ik}$ directly in this case, since now $A_{ij}$ depends
on them through the $F_i$.  One way to do it is by iteration, using,
say, the mean-field result (\ref{MFJ}) in (\ref{Fdef}) to get an initial
estimate of $F_i$, using this in (\ref{ATAP}) to get better $J_{ij}$s,
and so on. However in the stationary case this iteration can be
circumvented by a trick: Since we can write $J_{ik}$ as
$J_{ik}^{nMF}/(1-F_i)$, we can use (\ref{Fdef}) to obtain \be F_i =
\frac{(1-m_i^2)\sum_k (J_{ik}^{nMF})^2 (1-m_k^2)}{(1-F_i)^2},
			\lb{Fint} \ee which is a cubic equation that we can solve
for $F_i$, given the mean-field $\sf J$.  The relevant root is the one
that goes to zero as $J_{ij} \to 0$.  This root cannot exceed $1/3$,
which restricts this method to fairly weak couplings.  Still, it is an
improvement on the mean-field approximation.

In the same way as for the MF case, once the $J_{ij}$ are found, we can
go back to the TAP equations in the form (\ref{TAPhofm}) to obtain the
$h_i$.

\vspace{0.5 cm} \noindent {\bf 3.3  Nonstationary case:  exact algorithm
} \vspace{0.2 cm}

Most of the above carries over to the nonstationary case, where the
$h_i$ depend on $t$.  However, now different time bins are not
statistically equivalent, so we require data from (many) runs of the
system.  Thus, now we will write the spike trains as $S_i(t,r)$, where
$r$ labels the different runs and $t$ is the time index within a run. 
The ``magnetizations'' are now $t$--dependent averages over runs, \be
m_i(t) = \langle S_i(t,r) \rangle_r,				\lb{m(t)def} \ee and
we define $\delta S_i(t,r) = S_i(t,r) - m_i(t)$.

In (\ref{LL})  $S_i(t+1)$ and $H_i(t)$ acquire a run argument $r$.  The
learning rules analogous to (\ref{db}) and (\ref{dJ2}) are again
straightforward to derive by differentiating with respect to $b_i(t)$
and $J_{ij}$: \bea \delta b_i(t)  &=& \eta \left [ m_i(t+1)  - \left
\langle \tanh H_i(t,r) \right\rangle_r \right]     \lb{dhnonstat} \\
\delta J_{ij} &=& \eta T \left [  D_{ij} -  \left \langle \left [\tanh 
H_i(t,r) - m_i(t+1)\right] \delta S_j(t,r)\right
\rangle_{r,t}	\right],		\lb{dJnonstat} \eea where $T$ is the
run length and \be D_{ij} = \langle \delta S_i(t+1,r) \delta S_j(t,r)
\rangle_{r,t}.  \lb{Dnsdef} \ee Note that it is not the same as the
$D_{ij}$ (\ref{Ddef}) because now the $\delta S_i$ are defined relative
to the time-dependent run averages $m_i(t)$, not the time-independent
$m_i$.

When we use the word ``exact" to describe this algorithm, we mean it in
the same sense as for the stationary algorithm, but ``infinite data''
now means an infinite number of runs to average over in computing the
statistics.

\vspace{0.5 cm} \noindent {\bf 3.4  Nonstationary case:  mean-field
methods } \vspace{0.2 cm}

First one needs time-dependent mean-field and TAP equations.  These have
been derived recently (Roudi and Hertz, 2011b), using a systematic
scheme analogous to that employed by Plefka for the equilibrium case. 
The dynamic TAP equations for the time-dependent magnetizations take the
form \be m_i(t+1) = \tanh\left[ h_i(t) + \sum_j J_{ij}m_j(t)
-m_i(t+1)\sum_j J_{ij}^2 (1-m_j^2(t))\right] \lb{dynTAP} \ee For (naive)
mean field, the equations are the same except that the last term inside
the brackets is absent.

Now the same procedure that led to (\ref{MF2}) gives (Roudi and Hertz,
2011a) \be D_{ij} = \sum_k J_{ik}\langle
(1-m_i^2(t+1))C_{kj}(t)\rangle_t,			\lb{nsMF} \ee where \be
C_{kj}(t) = \langle \delta S_k(t,r) \delta
S_j(t,r)\rangle_r.				\lb{Coft} \ee If we define a set of
matrices ${\sf B}^{(i)}$ by \be B^{(i)}_{kj} = \langle
(1-m_i^2(t+1))C_{kj}(t)\rangle_t,				\lb{defB} \ee we can
again solve for $\sf J$ by simple matrix algebra: \be J_{ij} = \sum_k
D_{ik} [({\sf
B}^{(i)})^{-1}]_{kj}.						\lb{JMFnonstat} \ee The
calculation is a little more time-consuming than the stationary one,
because now we have to invert a  different matrix ${\sf B}^{(i)}$ for
each $i$.  Nevertheless, it is still much faster than using the exact
algorithm.

For TAP, in the same way as in the stationary calculation, a correction
factor $1-F_i$ appears.  Now $F_i$ is time-dependent, \be F_i(t) =
(1-m_i^2(t+1))\sum_k J_{ik}^2 (1-m_k^2(t)),		  \lb{Fnonstatdef}
\ee and the $(1-F_i)$-factor enters the time average that defines the
matrices ${\sf B}^{(i)}$: 
\be B^{(i)}_{kj} = \langle(1-m_i^2(t+1))(1-F_i(t))C_{kj}(t)\rangle_t,				
\lb{defBnonstat} \ee 
Now there is no simple equation one can solve for
$F_i(t)$, as there was in the stationary case, so solving for the $\sf
J$-matrix has to be done iteratively.

However, we have found that qualitatively good reconstruction can be
done if we make the approximation of factoring the average in
(\ref{defBnonstat}): \be \langle
(1-m_i^2(t+1))(1-F_i(t))C_{kj}(t)\rangle_t \approx (1-F_i)\langle
(1-m_i^2(t+1))C_{kj}(t)\rangle_t,					\lb{factorization}
\ee where $F_i = \langle F_i(t)\rangle_t$.  Now $F_i$ solves a cubic
equation, as in the stationary case: \be F_i (1-F_i^2)= \sum_k
(J_{ik}^{nMF})^2\langle (1-m_i^2(t+1))
(1-m_k^2(t))\rangle_t,			\lb{Fintapproxns} \ee and, given
$F_i$, one has $J_{ij} = J_{ij}^{nMF}/(1-F_i)$, again as in the
stationary case.

\vspace{0.75 cm} \noindent {\bf \large 4  Errors } \vspace{0.2 cm}

In a Bayesian way, we can interpret the likelihood $\exp L({\sf J},{\bf
b})$ as a probability density for the $J$s and $b$s with a flat prior. 
This density will peak around the parameters found by the algorithm, and
if the data are generated by the kinetic Ising model itself, the peak
will lie at their true values in the limit of infinite data.  The spread
of the probability density around the peak  gives a measure of how the
parameters obtained from a single data set will vary from one data set
to another.  Because the log likelihood is proportional to the data set
size, it is sufficient, for large $T$ (stationary case) or many runs
(nonstationary case) to expand it to second order in the deviations
$\delta J_{ij}$ from the values ${\sf J}_0, {\bf b}_0$ at the maximum,
so the resulting density is a Gaussian.  Suppressing, for simplicity,
the dependence on the $b$s and the overall normalization, we find, for
the nonstationary case, \be P({\sf J}) \propto \prod_i \exp \left(
L({\sf J}_0,{\bf b}_0)- \half  RT \sum_{jk} B_{jk}^{(i)}\delta
J_{ij}\delta J_{ik}\right), \lb{GaussianP} \ee where $R$ is the number
of runs and the $B_{jk}^{(i)}$ are the matrix elements that occurred
above (\ref{defBnonstat}) in the mean-field algorithm.  Here we can see
that  $RTB_{jk}^{(i)}$ are the elements of the Fisher information matrix
of the distribution of $J$s.  The factor of $RT$ means that the
fluctuations of the $J$s around their central values go to zero  like
$1/\sqrt{RT}$ for large data sets.  This makes more precise our
statement that the exact algorithms described above recover the correct
model parameters correctly in the limit of infinite data, if those data
are generated by our kinetic Ising model.

To get a little more insight, consider the stationary case, where
$B_{jk}^{(i)} = T(1-m_i^2)C_{jk}$.  The one can see (1) that $J_{ij}$s
with different first (postsynaptic) indices are uncorrelated, and (2)
$J_{ij}$s with different second (presynaptic) indices will have strongly
correlated errors if they have large projections along eigenvectors of
$\sf C$ that have small eigenvalues.   Thus, elements of $\sf J$
identified as large by the mean field approximation (\ref{MFJ}) are
exactly the ones for which the estimate is most uncertain.  Finally, if
correlations in the data are very weak, $\sf C$ is almost diagonal, and
\be \langle \delta J_{ij}\delta J_{ik} \rangle =
\frac{\delta_{jk}}{T(1-m_i^2)(1-m_j^2)}.	\lb{diagapp} \ee These
considerations show that reconstruction is harder when the data are
strongly correlated and when rates are low ($m_i$ near -1).

For the mean-field or TAP algorithms, the distribution of $J$s will have
a form like  (\ref{GaussianP}), but it will not be centered at the true
maximum.  The mean square errors will be \be \langle \delta J_{ij}\delta
J_{ik} \rangle= \frac{[({\sf B}^{(i)})^{-1}]_{jk}}{RT}  +  (J_{ij}^0 -
J_{ij}^{\rm true}) (J_{ik}^0 - J_{ik}^{\rm true})
.							\lb{approxerr} \ee In cases (such as weak
coupling) where the approximations are good, the mean square errors will
show the $1/RT$-dependence for small data sets, but for large $RT$ they
will level off at a residual value given by the second term in
(\ref{approxerr}).

If the data come from a different model than our kinetic Ising one, we
can never obtain its true parameters, even if we use the ``exact''
algorithms.   In this case we say that the model that generated the data
was ``unrealizable''.  The only ``realizable'' cases are ones where the
model used to make the fit is the same as the one used to make the data.
 When the two models are in some sense close to each other and the $J$s
have a meaning in the true model (for example, if we added a term  with
weak delayed interactions $\sum_j K_{ij}S_j(t-1)$ to (\ref{fieldatt})),
we can have a situation like that for the approximate algorithms, with
errors that first decrease with data set size and then level off at
residual values.  But in general all we can say is that we have found
the best kinetic Ising model, i.e., the kinetic Ising model most likely
to have generated the data.

The real ``true'' model, i.e., nature itself, is far from our kinetic
Ising model.  Neurons have much more complex dynamics than our model
endows them with, and synapses have dynamics of their own that we have
ignored entirely in the present formulation.  Thus, in applying our
methods to neural data, we have to be rather cautious about interpreting
the connection strengths $J_{ij}$ that we obtain.

Furthermore, even if we could replace our simple binary units by models
with realistic dynamics and include suitable models for the synapses, we
could not expect to recover their parameters correctly from an inference
procedure like this, because most of the neurons in the true network are
not recorded.  A typical experiment records spikes of about 100 neurons
in a region of cortex of area about 10 ${\rm mm}^2$, which contains of
the order of a million neurons.  Hence, even ignoring the fact that
neurons in this region are coupled to many outside it, we have data from
only about 0.01\% of the local network.  It is clearly an audacious
project to hope to learn about the network connections and dynamics from
such limited data (hence the resort to terms like ``functional
connectivity'' to describe the results).

On the other hand, we have to start somewhere, and the simple
description of a neuron as a stochastic element that fires at a rate
that is a sigmoidal function of its net input is at least qualitatively
consistent with neurophysiology.  While we cannot expect to associate
all the $J_{ij}$s we find with direct synaptic connections, we would be
surprised if the strongest $J_{ij}$s we found were not indicative of
synapses between recorded neurons (assuming we had chosen the time bin
size for our data sensibly, i.e., around the sum of typical synaptic and
membrane time constants).  Our findings (Roudi and Hertz, 2011a; Sect
10.6.2 below), as well as results obtained applying this approach to
genetic regulatory network networks (Lezon et al, 2006), support this
optimism.

\vspace{0.75 cm} \noindent {\bf \large 5 Regularization } \vspace{0.2
cm}

The algorithms above can be given a simple little extra twist that
allows us to eliminate the smallest (presumably, least significant)
connections in the network in a controlled way.   The idea (Tibshirani,
1996; Ravikumar et al, 2010) is to try to minimize a cost function \be E
= -L[{\bf S}, {\sf J}, {\bf h}] +\lambda \sum_{ij} | J_{ij}
|.			\lb{cost} \ee In a Bayesian perspective, the addition of
this term is the imposition of a prior $\exp(-\lambda \sum |J_{ij}|)$ on
the $J_{ij}$.  It leads to a new term in the update rules (\ref{dJ}),
(\ref{dJ2}) and (\ref{dJnonstat}) for $J_{ij}$ proportional to $-\lambda
 {\rm sgn} J_{ij}$.   To see its effect, think of the space of $J$s,
with an axis for each pair $ij$. Under learning, the extra term pushes
the $J$ vector toward the edges of a hyper-octahedron where many of its
components vanish.  The regularization parameter $\lambda$ is chosen
small, so this is a weak effect for large $J_{ij}$.  But it is a big
effect for $J_{ij} \sim \lambda$ or smaller, and it tends to drive these
$J$s to zero.  Thus this modification of the learning performs a kind of
automatic pruning of the network, the degree of which we can control by
adjusting $\lambda$.

A possibility for another kind of regularization arises in using the
nonstationary algorithms described above.   The source of the time
dependence of the driving fields $h_i(t)$ or $b_i(t)$ is the time
dependence of the measured $m_i(t)$, i.e., of the firing rates.  For
realistic amounts of data (perhaps 100 repetitions of the stimulus in an
experiment), one cannot expect to estimate a rate within a time bin to
better than 10\% accuracy.  The apparent rates will fluctuate this much
from one time bin to the next, and the inferred $h_i(t)$ will inherit
these fluctuations.  If we believe that the true rates vary more
smoothly than this, we can add a further term to the cost function, \be
E_{smooth} = \half \kappa \sum_{it}[h_i(t) -
h_i(t+1)]^2.			\lb{smoothcost} \ee The resulting extra term in
$\delta h_i(t)$ is proportional to $\kappa [h_i(t-1)-2h_i(t)+h(t+1)]$.
This smoothes $h_i$ by averaging $h_i(t)$ a little with $h_i(t\pm 1)$.

\vspace{0.75 cm} \noindent {\bf \large 6  Testing the Models}
\vspace{0.2 cm}

We have tested these models on three kinds of data:

\noindent (1) Data generated by the models themselves, where we can
measure directly how well they recover the (known) network parameters.

\noindent (2) Data generated by a semi-realistic computational model of
a cortical network.  This model is much more complex than the kinetic
Ising model we try to fit its data with, but we still know the true
connectivity of the network.

\noindent (3) Data recorded from a real biological network, in this case
of salamander retinal ganglion cells.

\vspace{0.5 cm} \noindent {\bf  6.1  Testing the models on themselves}
\vspace{0.2 cm}

Testing the algorithms described above on how well they reconstruct the
couplings of a kinetic Ising model has two purposes: (1) For the exact
algorithms, to verify that they work according to the theoretical
picture set out above, in particular, checking that the mean square
errors fall off like $1/T$ (stationary case) or $(RT)^{-1}$
(nonstationary case).  (2) For the approximate algorithms, evaluating
the average residual error in (\ref{approxerr}) and studying how it
depends on the strength of the couplings in the model.  We summarize the
findings  (Roudi and Hertz, 2011a) here.

\begin{figure}[h]
\centering 
\subfigure[]{\includegraphics[height=6 cm,width=6 cm]{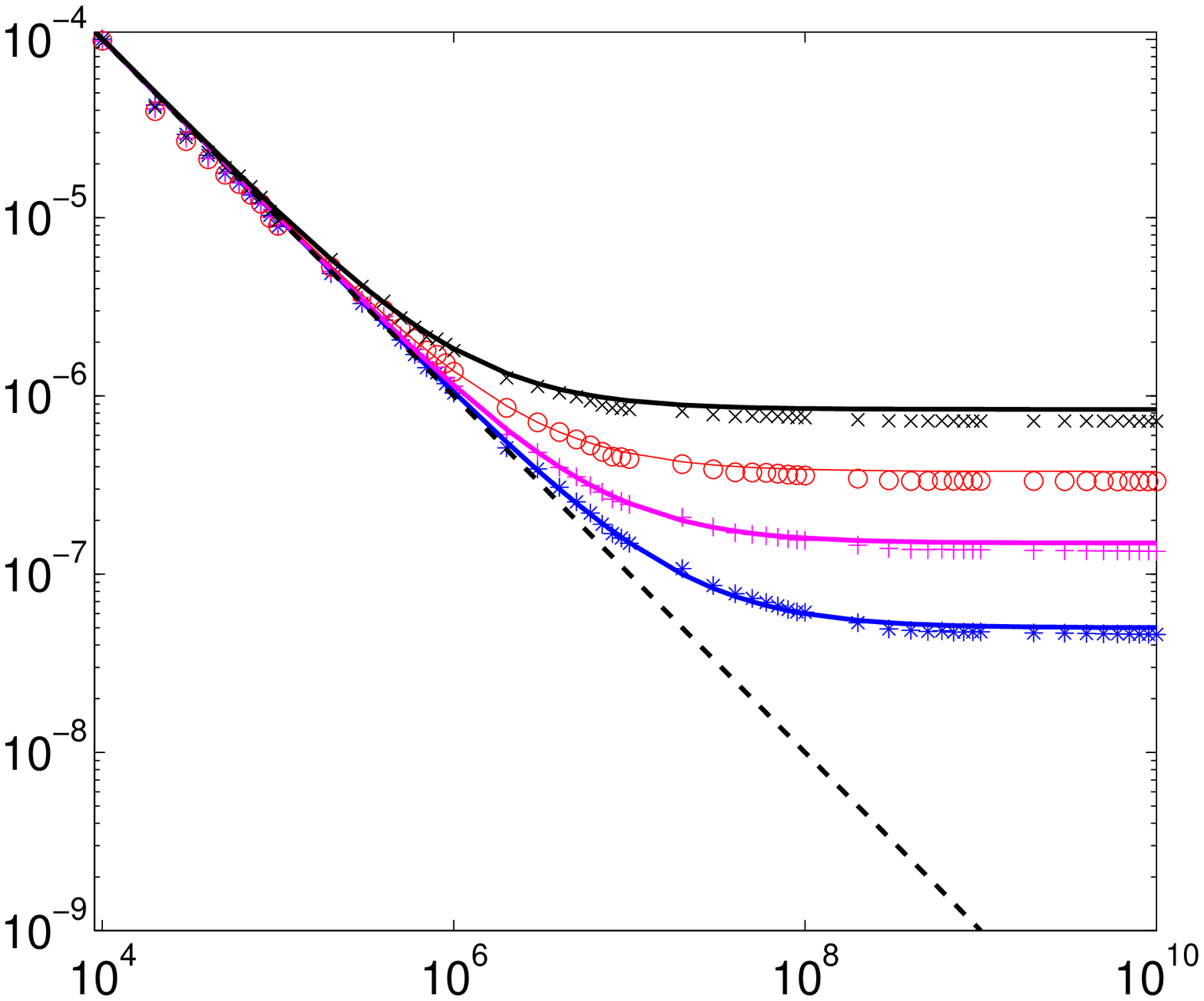}} 
\subfigure[]{\includegraphics[height=6 cm,width=6 cm]{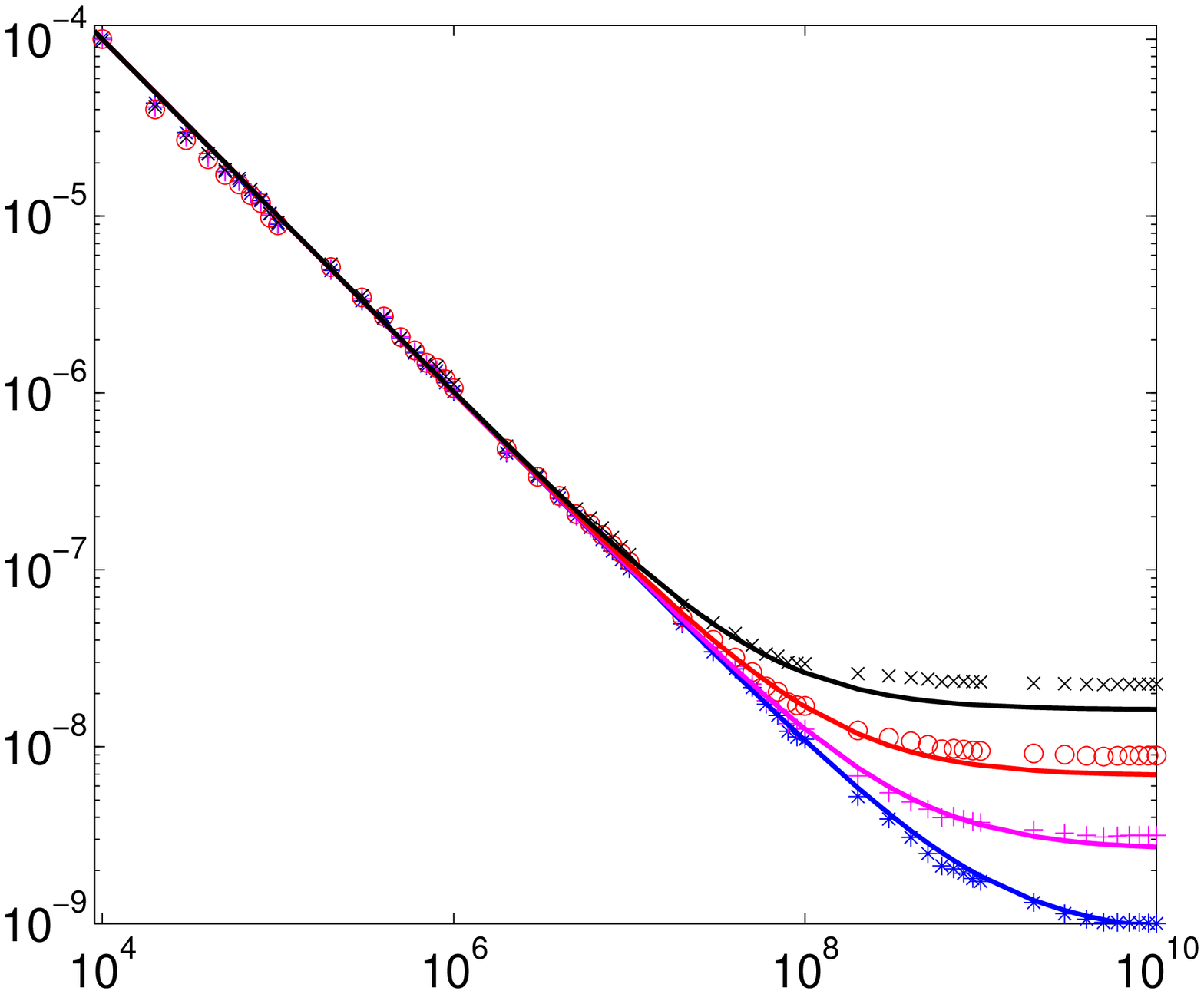}} 
\caption{Mean square errors for nMF (a) and TAP
(b) approximations, stationary kinetic Ising model, as a function of the
number of time steps $T$  in the training data for noise levels g = 0.1
(*), 0.12 (+), 0.14 (circles), and 0.16 (x).  The solid lines are the
theoretical predictions from equations (52) (nMF) and (53) with
finite-size corrections (TAP), and the symbols are the errors measured
for the $J$s obtained from the algorithms. The dashed line is the rms
error $1/T$ for the exact algorithm. (adapted from Roudi and Hertz,
2011)} 
\label{Fig1} 
\end{figure}

Fig. 1 shows mean square errors in the $J$s for a model in which the
$J_{ij}$ are chosen independently from a Gaussian distribution of mean
zero and standard deviation $g/\sqrt{N}$, where $N$ is the number of
neurons in the network.  Here we have taken $N = 20$ and all $h_i = 0$. 
The nMF and TAP errors follow the exact-algorithm result $1/T$ for small
$T$ and then flatten out at minimum values (higher for nMF that TAP).

The asymptotic nMF and TAP errors are systematic: For infinite data, nMF
systematically underestimates the magnitude of the $J_{ij}$ and TAP
overestimates them (though less so).  In fact, the factor $1-F_i$
relating the TAP and nMF $J_{ij}$s is the lowest-order correction of the
nMF underestimation.  We can see this simply, for zero field, by
expanding the tanh in (\ref{dJ}) or (\ref{dJ2}) at $\delta J_{ij} = 0$
to third order in the $J$s (instead of just to linear order as we did in
(\ref{MF1})). We get \be D_{ij} = \sum_k J_{ik} C_{kj} - \third
\sum_{klm}J_{ik}J_{il}J_{im}\langle S_kS_lS_mS_j\rangle + \cdots
.										\lb{mfexpand} \ee (All
correlations here are equal-time.)  In the sum over $k$, $l$ and $m$,
terms with $k=l$, $l=m$ and $m=k$ dominate.  Multiplying on the right by
the inverse of ${\sf C}$ and using the nMF result (\ref{MFJ}), we find
\be J_{ij}^{\rm nMF} = J_{ij} - \left( \sum_k J_{ik}^2 \right)
J_{ij}	 			\lb{JnMFerr} \ee (plus corrections of relative
order $1/N$).  Now we can use the fact that for our SK-like model, the
quantity $\sum_k J_{ik}^2 = g^2$, again ignoring corrections of higher
order in $1/N$, so we obtain \be J_{ij} = \frac{J_{ij}^{\rm
nMF}}{1-g^2}	.							\lb{JnMFcorrected} \ee The
denominator can be recognized as the TAP correction factor $1-F_i$ when
the $h_i=0$.   This implies an asymptotic mean square error \be
\epsilon_{\rm nMF} = \langle (J_{ij}-J_{ij}^{\rm nMF})^2\rangle =
\frac{g^6}{N}.		\lb{enMF} \ee The solid curves in Fig 1a are $1/T
+ \epsilon_{\rm nMF}$, evidently a quite good fit for the values of $g$
from 0.1 to 0.16 for which they are plotted.  The black dots in Fig 2
show the nMF $J$s plotted against the true ones.  When the size of the
data set is increased from $10^4$ to $10^6$, the scatter is reduced and
the systematic underestimation becomes clear.  In the limit of infinite
data, we expect the scatter to disappear, leaving a clean linear
behavior near $J_{ij}=0$ with a slope $1-g^2$.

\begin{figure}[h] 
\centering 
\subfigure[]{\includegraphics[height=6 cm,width=6 cm]{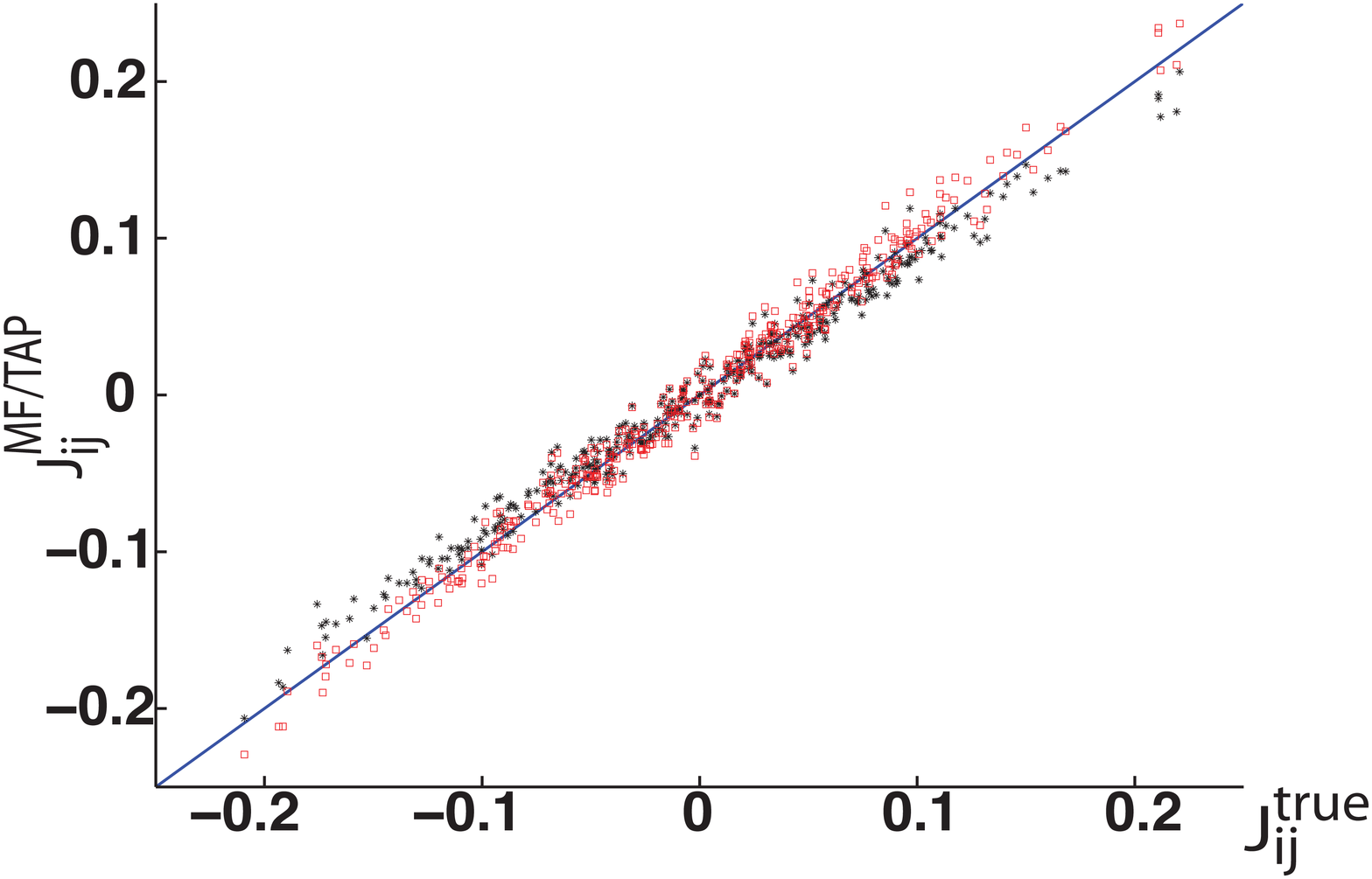}}
\subfigure[]{\includegraphics[height=6 cm,width=6 cm]{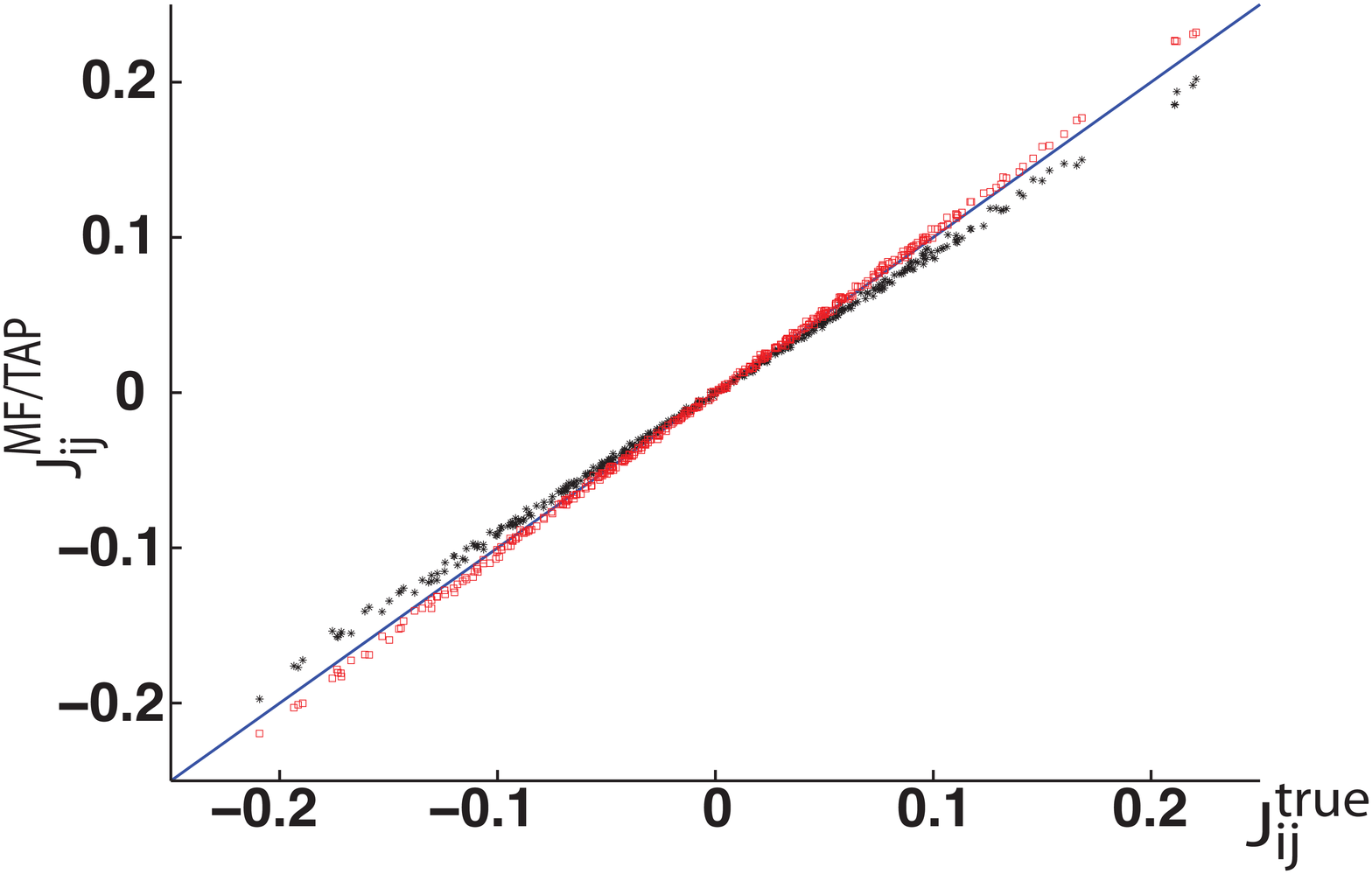}} 
\caption{$J$'s obtained by nMF (circles) and TAP
(crosses) algorithms, plotted against the true $J$s for a network of 20
neurons with $g = 0.35$. (a) $T = 10^4$; (b) $T=10^6$.} 
\label{Fig2}
\end{figure}

The TAP errors shown in Fig 1b are much smaller that the nMF ones,
because they include the lowest-order correction (\ref{JnMFcorrected}). 
We can find the leading error in the TAP approximation by expanding the
tanh in (\ref{dJ}) to fifth order.  A little algebra like that leading
to (\ref{enMF}) leads to an asymptotic error estimate \be \epsilon_{TAP}
= \frac{4g^{10}}{N}.							\lb{eTAP} \ee
Furthermore, one can show that TAP systematically overestimates the
magnitudes of the $J$s, in the same way that nMF underestimated them. 
For a large network, we expect this to describe the asymptotic TAP
error.  However, there are finite-size corrections, of order $g^6/N^3$. 
These would be negligible for $N \gg 1/g^2$,  but they are not for the
small network ($N=20$) at the values of $g$ studied here.  However,
taking both (\ref{eTAP}) and the finite-size corrections into account,
we are able to account for the measured errors (Fig. 1b).

In Fig 2 one can see that the TAP $J$s (square markers) lie much closer
to the true ones than the nMF $J$s (*) do, but a careful look at Fig 2b
reveals the systematic overestimation.

Finally, we give an example of nonstationary inference.  We generated $R
= 100$ runs, each of length $T= 10^5$ steps for a model with $h_i = 0.5
\cos (2\pi t/100)$.  The couplings are small enough ($g = 0.05$) that
nMF is quite accurate, as Fig 3a shows.  Furthermore, the fields are
also quite well recovered (dots, Fig 3c) by solving (\ref{dynTAP})
(without the TAP term) for $h_i(t)$.

\begin{figure}[h] \centering 
\subfigure[]{\includegraphics[height=5 cm,width=5 cm]{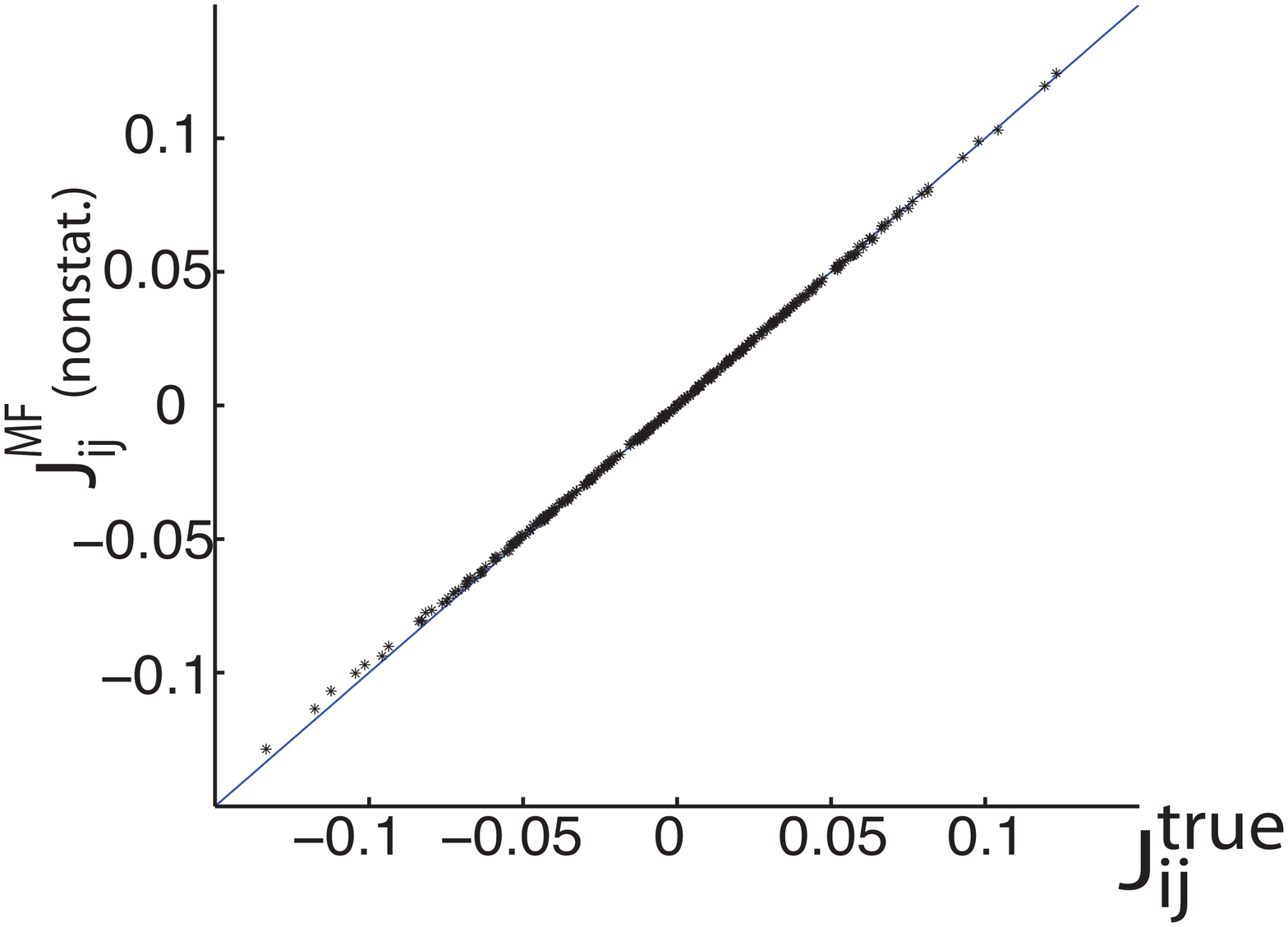}} 
\subfigure[]{\includegraphics[height=5 cm,width=5 cm]{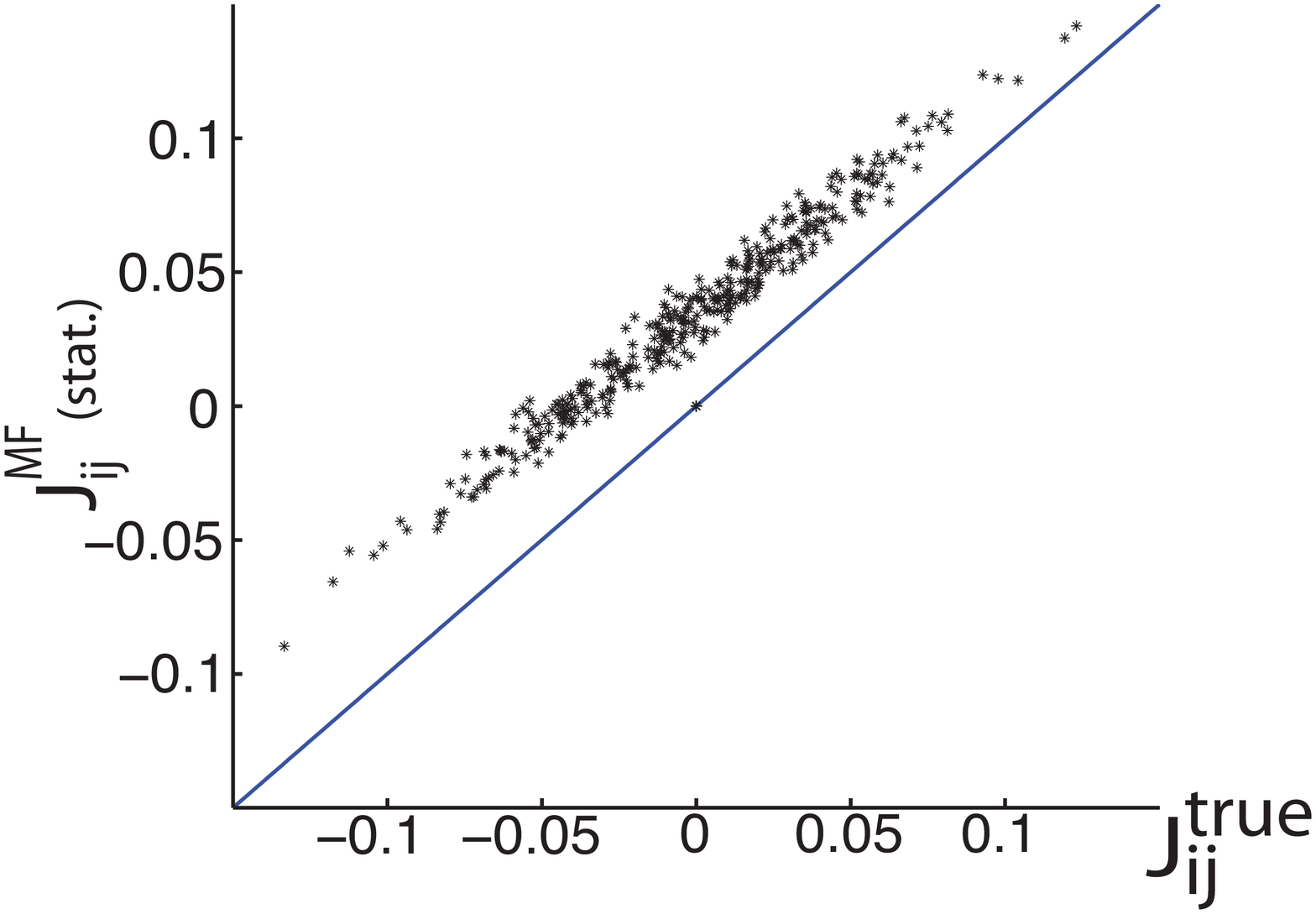}} 
\subfigure[]{\includegraphics[height=2 cm,width=6 cm]{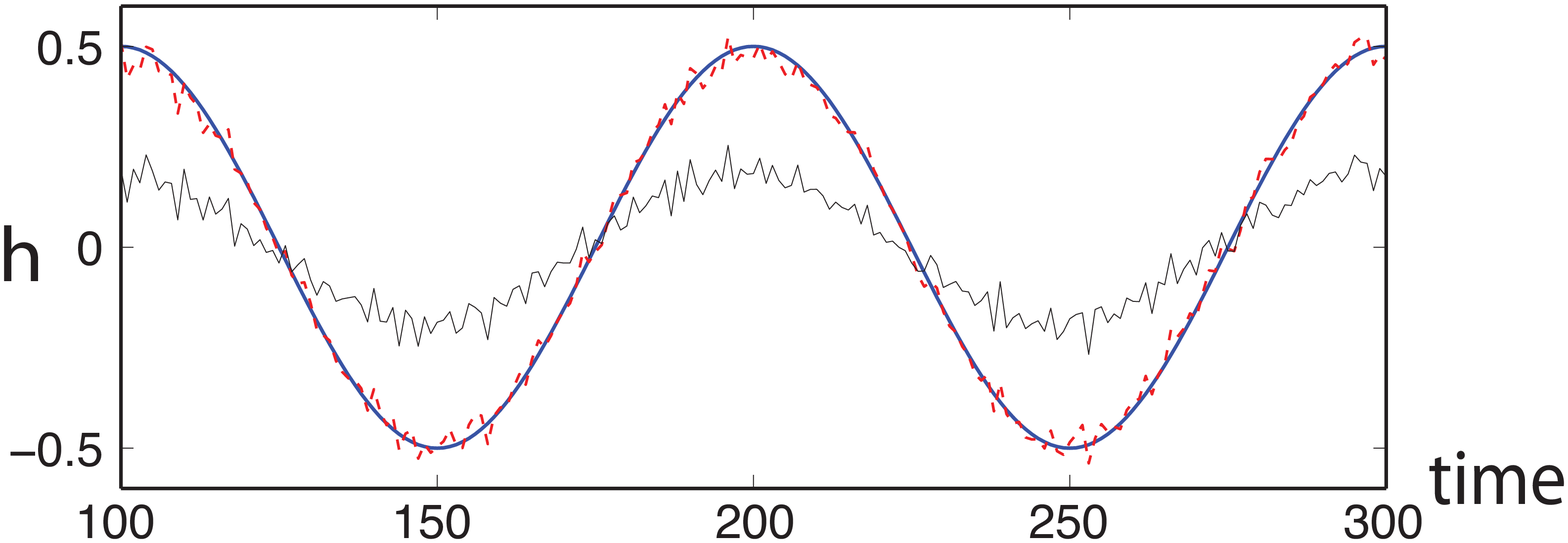}} 
\caption{Nonstationary inference: 20 neurons,
$g=0.05$, driven by a sinusoidal field with period 100 steps.  (a) The
couplings inferred using the nonstationary nMF algorithm are almost
equal to the true ones.   (b) Couplings inferred using the stationary
algorithm are systematically overestimated. (c) The nonstationary
algorithm recovers the driving field (solid curve) correctly (dots),
while the amplitude of the field predicted from the mean field equations
using the couplings found by the stationary algorithm (crosses) is too
small. (adapted from Roudi and Hertz, 2011)}
\label{Fig3} \end{figure}

The $J$s obtained if one does the calculation assuming stationarity are
systematically too large (Fig 3b).  This is because of the apparent
correlations induced in the data by the common external field.  In
calculating correlations we should actually compute $\langle \delta
S_i(t) \delta S_j(t') \rangle_r$ using $\delta S_i(t) = S_i(t) -
m_i(t)$, with time-dependent $m_i(t)$.  However, if we assume
stationarity and use $\delta S_i(t) = S_i(t) - \langle m_i(t)
\rangle_t$, instead, we will overestimate the correlations.  (This is
sometimes called ``stimulus-induced correlations''.)  These spurious
correlations  (all positive, in this case) then lead to a spurious
increase in the inferred $J_{ij}$s.   One can use these spurious $J$s in
(\ref{dynTAP}) (again, here with the TAP term absent) to infer $h_i(t)$,
but because the $J$s are too big, the resulting $h_i(t)$ do not have a
modulation amplitude as large as they should (Fig 3c, crosses).

\vspace{0.5 cm} \noindent {\bf  6.2  Application To Data From Model
Cortical Network} \vspace{0.2 cm}

For a first nontrivial test of the approach, we try it on data generated
by a realistic model cortical network (Hertz, 2010).  This is a network
of 1000 spiking neurons, 80\% excitatory and 20\% inhibitory, with
Hodgkin-Huxley-like intrinsic dynamics.  They are driven by an
additional external population of 800 Poisson neurons firing tonically
at 10 Hz.  All neurons, both internal and external, are connected by
conductance-based synapses in a random fashion, with a connection
probability of 10\% (except that the external neurons are not connected
to each other).    The strengths of these synaptic conductances were
chosen so that the network was in a high-conductance balanced state
(Amit and Brunel, 1997; van Vreeswijk and Sompolinsky, 1998) with
average conductances in the range measured experimentally (Destexhe et
al, 2003).

The magnitudes of the synaptic conductances did not vary within a class
(excitatory-to-excitatory, excitatory-to-inhibitory, etc.), except for
the fact that 90\% of them were zero because of the random dilution,
though they did vary somewhat in their temporal characteristics.  For
such a system, a natural and important question to ask is how well we
can identify the connections that are actually present.

We choose for analysis the spike trains of the inhibitory neurons with
rates over 10 Hz.   There were 95 of them.  Their mean rate was 32 Hz,
with a maximum of 83 Hz.  Higher firing rates give better statistics,
and the inhibitory synapses are the strongest ones in this model, so
this choice makes the task of fitting the model and identifying the
connections present easier than it might otherwise be.  However, it is
useful to test the method on an easier problem before trying to solve a
more difficult one.  The data were put into the form to be used by the
algorithms by binning the spikes with 10-ms bins.

To study how much data is necessary for the fit, we did the analysis
using the exact stationary algorithm for data sets of size $T$ from 1000
up to 64000 time bins. For each such $T$, we calculated the log
likelihood of the data.  For comparison, we also calculated these
numbers for independent-neuron models based on the same partial data
sets.  The results were adjusted with Akaike corrections equal to the
number of parameters in the models (Akaike, 1974).

Fig 4 shows the log likelihoods (per time step, per neuron), with and
without Akaike corrections.  They appear to converge to a value of
$-0.306 \pm 0.001$, and evidently there is little point in using more
than around 50000 time steps of data.  Also shown are the log
likelihoods for an independent-neuron model (all $J_{ij} = 0$).  For
these, the Akaike correction is so small that the corrected and
uncorrected curves would fall almost on top of each other, so only the
adjusted values are actually plotted.  The curve is nearly flat at at
value of about -0.53.  It is evident that the model with $J$s is much
better than the one without them.

\begin{figure}[h]
\centering 
\includegraphics[height=7 cm, width=7cm]{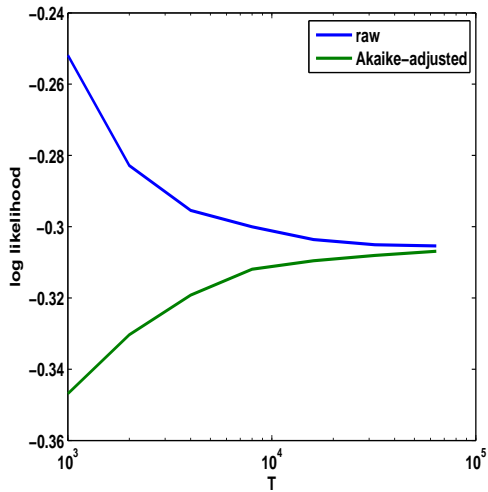} 
\caption{Cortical model data (tonic firing):  Log
likelihoods as a function of data set size $T$ for fits with kinetic
Ising and independent-neuron models.  Dashed line:  log likelihood.
Solid line: Akaike-adjusted log likelihood.  Dotted line:
independent-neuron model ($J_{ij}=0$). Despite its greater number of
parameters, the model with couplings is clearly better, and the
difference is evident for all the data set lengths in the range shown.}
\label{Fig4} 
\end{figure}

The Akaike-adjusted log likelihood is a suitable statistic for comparing
model quality, but one does not have an clear idea of what a value of
-0.306 means for the quality of the network reconstruction, in
particular for the problem of identifying the connections present in
this diluted network.  To do this, we use another statistic, defined as
follows.

\begin{figure}[h] 
\centering 
\includegraphics[height=7 cm, width=7cm]{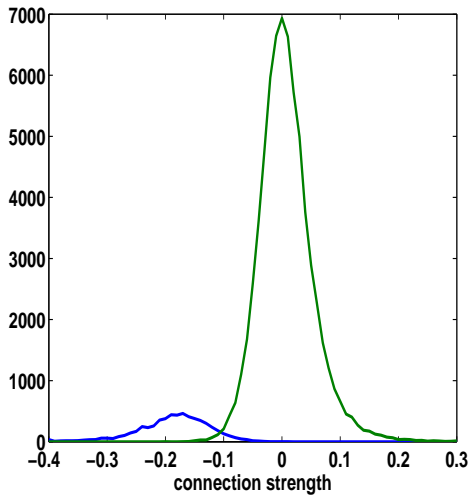} 
\caption{Distributions of connection strengths found for
cortical model network data for inhibitory connections present in the
network (solid line) and connections not present in the network (dashed
line).  Averages of results of performing nMF inference on 30 sets of 50
neurons chosen randomly from the 95 inhibitory neurons with rates $> 10$
Hz. } 
\label{Fig5} 
\end{figure}

Consider the values of $J_{ij}$ that the algorithm finds for the
connections that are actually present in the network.  These estimates
will have some spread around their mean value, because of the limited
data and the mismatch between the kinetic Ising model and the real
network.  For the same reason, the values assigned to the $J_{ij}$s  for
which the connections $\{ ij\}$ are not present, which should be zero,
will also be spread around their (small) mean value (Fig 5).  If the
spreads of these two distribution are small compared to the difference
between their means, we can easily identify the true connections, even
if we do not know which ones they are {\em a priori}.  A measure of the
difficulty of the task which we can compute (knowing the true
connectivity) is the noise/signal ratio \be d \equiv \frac{\sigma(J_{\rm
true} \neq 0)+\sigma(J_{\rm true}=0)}{|\langle J \rangle_{J_{\rm true}
\neq 0} - \langle J \rangle_{J_{\rm true} = 0}| }, \ee where the
standard deviations and means are over the $J$s found by the algorithm.
Fig 6 shows $d$ as a function of data set size $T$, calculated for the
$J$s obtained by both the exact algorithm and the nMF and TAP
approximate formulae.    The TAP and nMF results are too close to each
other to distinguish, so only the latter is plotted.  The plot in Fig 5
was made using TAP results for $T=200000$ and shows the kind of errors
one makes for $d \approx 0.47$:  The false positive rate is 5.6\% and
the false negative rate is 7.2\%.  For $d \approx 0.3$, which one
achieves with the exact algorithm, there are almost never any errors.

\begin{figure}[h] 
\centering 
\includegraphics[height=7 cm, width=7cm]{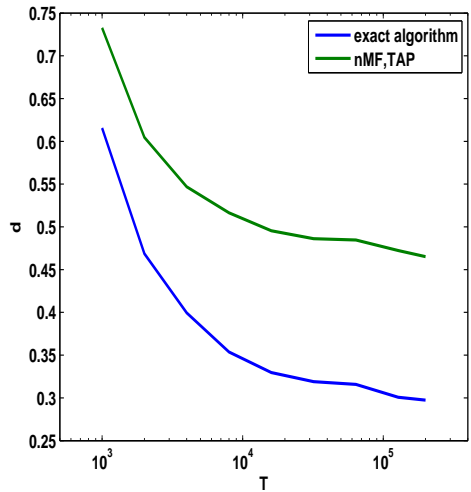} 
\caption{Fitting cortical network data: Noise/signal ratio
$d$, as function of data set length $T$ for nMF (dashed) and exact
(solid) learning algorithms.} 
\label{Fig6} 
\end{figure}

Note, furthermore, that this reconstruction was achieved for a strongly
undersampled network -- There were 1000 neurons in the network but the
reconstructions were done using only 50 of them at a time.  This lends
some support to our hope that reconstruction, at least for strong
synapses, might be possible even though we cannot record from all the
neurons in the network.

In this example, we used the stationary versions of the algorithms
because the rate of the external driving population was constant.  We
have also studied a case where the external rate varies in time,
performing the corresponding nonstationary analysis, which gives similar
results.  Results will be published elsewhere.

\vspace{0.5 cm} \noindent {\bf  6.3  Application To Data From Salamander
Retina} \vspace{0.2 cm}

We have also analyzed a data set provided by Michael Berry of recordings
from 40 ganglion cells in a salamander retina.  The retina was
stimulated 120 times by a 26.5-s movie clip (a total recording time of
3180 s).  A nonstationary treatment is natural, in view of the
time-dependent stimulus.  We expect a significant contribution to the
connections identified in a stationary analysis from stimulus-induced
correlations, as in Fig 3b.

Such data have been analyzed in the past using the Gibbs equilibrium
analysis described in Section 2 (Schneidman et al 2006), which assumes
stationarity.  It is interesting to see whether the $J_{ij}$s  obtained
that way are related to those obtained using a stationary kinetic Ising
model.   They are not expected to be the exactly the same, since the
Gibbs $J$s are determined solely by the equal-time correlations and the
kinetic Ising ones depend on the one-time-step delayed correlations. 
Nevertheless, we can hope that if we have chosen the time bin size
sensibly, the connections identified by the two methods will be similar.

To find out, we fit both the Gibbs model and the stationary version of
our kinetic Ising model to these data.  Fig. 7a shows a scatter plot of
the resulting two sets of $J_{ij}$s.  While they are not so similar as
to be concentrated along a straight line in the plot, they do tend to
have the same sign (most of the points are in the first or third
quadrants).

\begin{figure}[h] 
\centering 
\subfigure[]{\includegraphics[height=6 cm,width=6 cm]{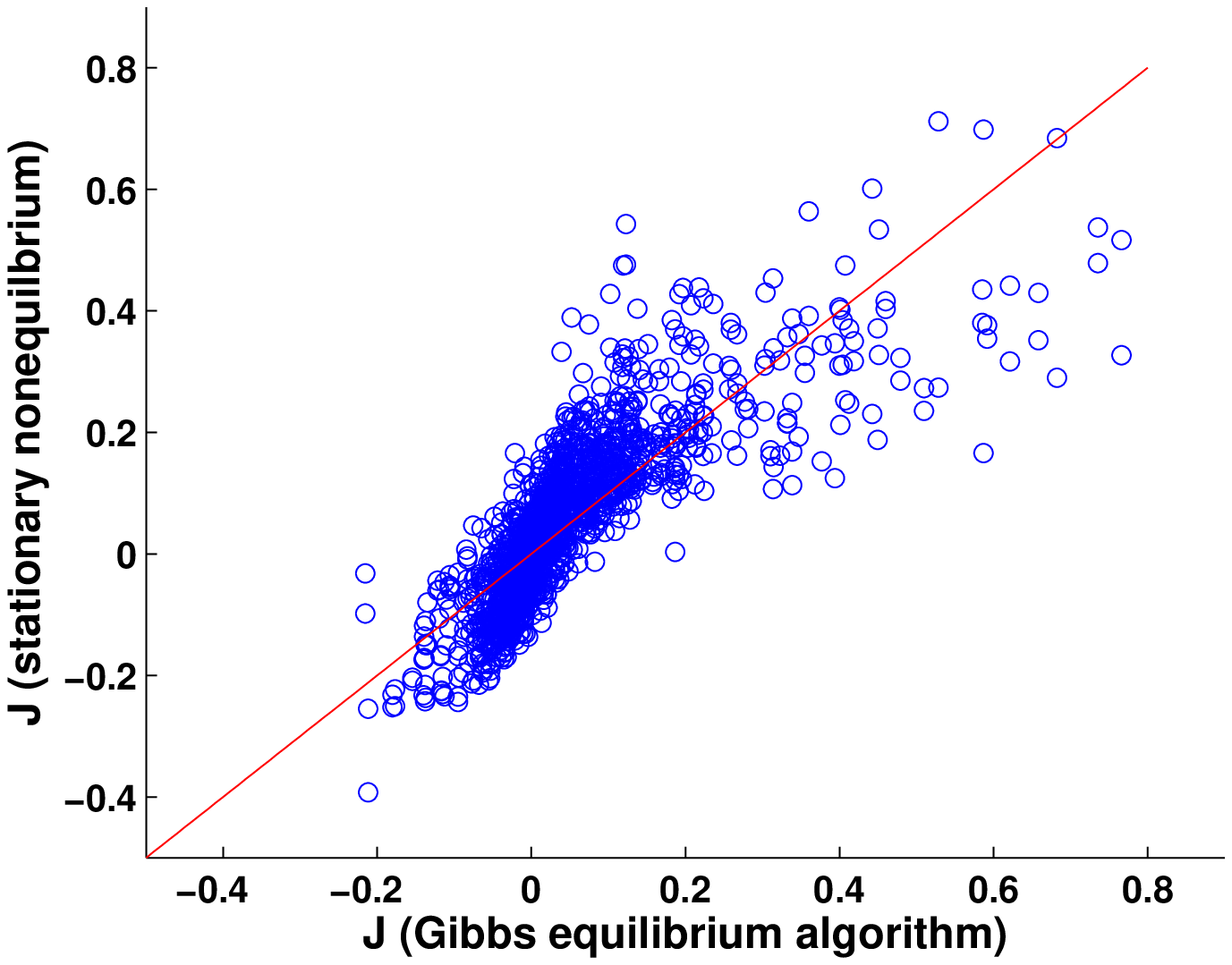}} 
\subfigure[]{\includegraphics[height=6 cm, width=6 cm]{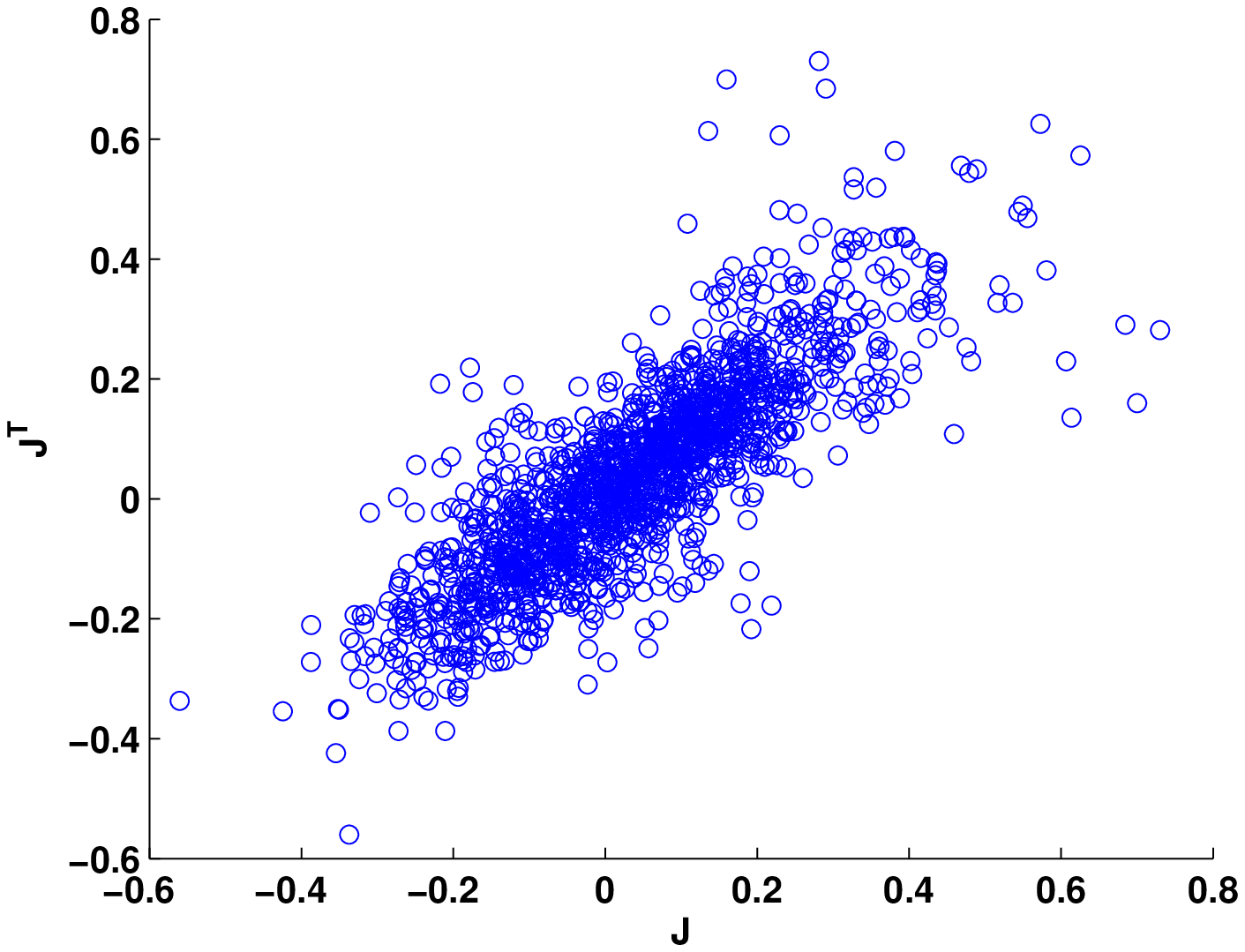}}
\caption{Fig 7. Salamander retina data: (a)
Couplings found from the stationary kinetic Ising model algorithm 
plotted against those obtained from the Gibbs equilibrium algorithm. (b)
Testing for the degree of symmetry in couplings: elements of the
coupling matrix $\sf J$ plotted against those of its transpose.}
\label{Fig7} 
\end{figure}

The kinetic Ising model allows for an asymmetric $\sf J$ matrix, while
symmetry is enforced in the Gibbs model.  To see how asymmetric the
kinetic Ising $\sf J$ matrix is here, we made a scatter plot of the
elements of $\sf J$ against those of its transpose (Fig 7b).  
Evidently, the magnitudes of $J_{ij}$ and $J_{ji}$ can be different, but
they do not differ wildly, and they almost always agree in sign. This is
consistent with the fact that they agree qualitatively with the fully
symmetric $J$s found for the Gibbs model and with our expectation that
they may be the result, to a strong degree, of stimulus-induced
correlations, which are symmetric by definition.

To find out to what degree, we then performed a nonstationary analysis,
with time-dependent $b_i(t)$ in our kinetic Ising model.  In addition,
we made a fit for a model (also nonstationary) with independent neurons,
in which all time dependence of firing rates is explained in terms of
the time-dependent $b$s.

We compared these three models  on the basis of the log likelihoods of
the data under them, with Akaike corrections for the different numbers
of model parameters.  We found that the stationary model
(Akaike-adjusted log likelihood per neuron per time bin $-0.128$ bits)
was significantly worse than the nonstationary ones ($-0.0927$ bits for
the independent model and $-0.0906$ bits for the full model).  The
difference between the fits given by the nonstationary models with and
without $J_{ij}$s is quite small (0.0021 bits, only 2.4\% of the total
log likelihood).  A full description of this analysis will be published
elsewhere.

We conclude that essentially all the connections found for the
stationary and Gibbs models come from stimulus-induced correlations. 
Once we include time-dependent $b$s in the model, adding $J$s gives
almost no improvement in the fit.  Although it is disappointing not to
be able to identify any important intrinsic connections in this network,
the result illustrates how the nonstationary analysis can uncover
features of the system that stationary models (including the Gibbs
model) can not.

\vspace{0.75 cm} \noindent {\bf \large 7  Further Developments }
\vspace{0.2 cm}

In all the above, we restricted the treatment to the simplest kind of
kinetic Ising model, for pedagogical purposes.  It is not hard to extend
the model in various ways to bring it closer to neurobiological reality.
 The most obvious way is to modify the memory-less dynamics
(\ref{fieldatt}-\ref{KIdef}) by letting the firing probability at $t+1$
depend on spikes earlier than $t$: \be H_i(t) = h_i(t) +\sum_{j,s =
1}^{\tau}J_{ij}(s)S_j(t-s+1). \lb{kernel} \ee Thus, each connection in
the model is characterized by a temporal kernel $J_{ij}(s)$.  It
describes synaptic and membrane potential dynamics. It is
straightforward to derive exact and mean-field learning algorithms for
such a model.

Models of this kind, called generalized linear models (GLMs), have been
studied in recent years (Truccolo et al, 2004; Okatan et al, 2006). 
They were used (Pillow et al, 2008) in an extensive study of the
signaling by a population of 27 monkey retinal ganglion cells.  They
have also been used (Rebesco et al, 2010) to track changes in connection
strengths in a network in which synaptic plasticity was induced by
microstimulation.  Thus, the feasibility and utility of such solutions
of the inverse problem for real neural networks have been demonstrated. 
The simple cases described in this chapter can be thought of as ``poor
man's'' GLMs.  Since they have fewer parameters, they may be
particularly useful models when data are limited.  It may also be useful
to extend the mean-field methods discussed here to GLMs.

Another couple of directions in which further development of these
models would be useful are the following:

(1) As mentioned above, one never records from all the neurons in a
network. A systematic approach to the inference problem in the presence
of ``hidden units'' is needed.  This problem could usefully be explored
first for the simplest model (\ref{fieldatt}-\ref{KIdef}) in a
realizable case, to gain some knowledge about how to model the hidden
part of the population.

(2) Even the model with temporal synaptic kernels assumes
``current-based'' synapses, i.e., that a presynaptic spike causes a
particular current to flow in or out of the postsynaptic neuron,
independent of the state of that neuron.  Actually, what the presynaptic
spike causes is the stochastic opening and closing of a channel,
selective for particular ions, in the postsynaptic membrane.  The
synaptic current (or, more precisely, its average over many presynaptic
transmitter release events) is then the product of the conductance of
the channel if it were open, the time-dependent probability that it is
open, conditional on the presynaptic spike, and the difference between
the instantaneous postsynaptic membrane potential and the reversal
potential for that channel.  The difference between such ``conductance
based'' synapses and current-based ones may be quantitatively small for
excitatory synapses, since their reversal potentials are far above the
firing threshold, but they are not for inhibitory ones, whose reversal
potentials are not much lower than typical subthreshold membrane
potentials.  It would be desirable and interesting to extend the kind of
modeling here to include a little of this potentially relevant
biophysics, starting perhaps with the limiting case of ``shunting
inhibition''.

(3) Finally, we mention a new mean-field inversion algorithm (M{\'e}zard
and Sakallariou, 2011; Sakallariou et al, 2011) which reconstructs
kinetic Ising models exactly when the $\sf J$ matrix is completely
asymmetric.  It would be interesting to see how well it works on data
from more realistic models or experiments.


\noindent Ackley D, Hinton GE, Sejnowski TJ (1985)  A learning algorithm
for Boltzmann machines. Cognitive Science 9:147-169.\\ Akaike H (1974) A
new look at the statistical model identification. IEEE Trans Aut Control
19:716Ð723.\\ Amit DJ, Brunel N (1997) Model of global spontaneous
activity and local structured activity during delay periods in the
cerebral cortex. Cerebral Cortex 7:237-252.\\ Destexhe A, Rudolph M,
Par{\'e} D (2003) The high-conductances state of cortical neurons {\em
in vivo}. Nature Reviews Neurosci 4:739-751.\\ Glauber RJ (1963)
Time-dependent statistics of the Ising model. J Math Phys 4:294-307.\\
Hertz, J (2010) Cross-correlations in high-conductance states of a model
cortical network. Neural Comp 22:427-447.\\ Kappen HJ, Rodriguez FB
(1998) Efficient learning in Boltzmann machines using linear response
theory. Neural Comp 10:1137-1156.\\ Lezon TR, Banavar JR, Cieplak M,
Maritan A (2006) Using the principle of entropy maximization to infer
genetic interaction networks from gene expression patterns. Proc Nat
Acad Sci USA 103:19033Ð19038.\\ McCulloch WS, Pitts W (1943) A logical
calculus of ideas immanent in nervous activity. Bull Math Biophys
5:115-133.\\ M{\'e}zard M, Sakallariou J (2011) Exact mean field
inference in asymmetric kinetic Ising systems. arXiv:1103.3433.\\ Okatan
M, Wilson MA, Brown EN (2005) Analyzing functional connectivity using a
network likelihood model of ensemble neural spiking activity. Neural
Comp 17:1927-1961.\\ Peretto P (1984) Collective properties of neural
networks: a statistical physics approach. Biol Cybern 50:51-62.\\ Pillow
JW, Shlens J, Paninski L, Sher A, Litke AM, Chichilnisky EJ, Simoncelli
EP (2008) Spatio-temporal correlations and visual signalling in a
complete neuronal population. Nature 454:995-999.\\ Plefka T (1982)
Convergence conditions of the TAP equation for the infinite-ranged Ising
spin glass model. J Phys A 15:1971-1978. \\ Ravikumar P, Wainwright M,
Lafferty JD (2010). High-dimensional Ising model selection using
$\ell_1$-regularised logistic regression. Ann Stat 38:1287-1319.\\
Rebesco JM, Stevenson IH, K{\"o}rding KP, Solla SA (2010) Rewiring
neural interactions by micro-stimulation. Frontiers System Neurosci
4:39.\\ Roudi Y, Hertz J (2011a) Mean-field theory for nonequilibrium
network reconstruction. Phys Rev Lett 106:048702.\\ Roudi Y, Hertz J
(2011b) Dynamical TAP equations for non-equilibrium spin glasses. J Stat
Mech P03031.\\ Roudi Y, Tyrcha J, Hertz J (2009) The Ising model for
neural data: model quality and approximate methods for extracting
functional connectivity.  Phys Rev E 79:051915.\\ Sakellariou J, Roudi
Y, M{\'e}zard M, Hertz J (2011) Effect of coupling asymmetry on
mean-field solutions of direct and inverse Sherrington-Kirkpatrick
model. arXiv:1106.0452.\\ Schneidman E, Berry MJ II, Segev R,  Bialek W
(2006) Weak pairwise correlations imply strongly correlated network
states in a neural population. Nature 440:1007-1012.\\ Tanaka T (1998)
Mean-field theory of Boltzmann learning. Phys Rev E 58:2302-2310.\\
Thouless DJ, Anderson PW, Palmer RG (1977) Solution of `solvable model
of a spin glass'. Phil Mag 35:593-601.\\ Tibshirani R (1996) Regression
shrinkage and selection via the lasso. J Roy Stat Soc B 58:267-288.\\
Truccolo W, Eden UT, Fellows MR, Donoghue JP, Brown EN (2004) A point
process framework for relating neural spiking activity to spiking
history, neural ensemble, and extrinsic covariate effects. J
Neurophysiol 93:1074-1089.\\ Van Vreeswijk C, Sompolinsky H (1998)
Chaotic balanced state in a model of cortical circuits. Neural Comp
10:1321-1371.
\end{document}